\documentclass[referee]{cjaa}           

\usepackage{graphicx}                   
\input{epsf.sty}                        
\input{psfig.sty}                       

\begin{document}

   \title{Gamma-ray background: a review
}


   \author{T.M. Kneiske
      \inst{}\mailto{kneiske@physik.uni-dortmund.de}
     }
   \offprints{T.M. Kneiske}                   

   \institute{Universitaet Dortmund,
   Experimentelle Physik 5,
44221 Dortmund,              
Germany
                               }
             \email{kneiske@physik.uni-dortmund.de}

   \date{Received~~2001 month day; accepted~~2001~~month day}

   \abstract{The gamma-ray background is still a subject under great debate.
   All phenomena in the universe emitting gamma-rays can contribute directly
   as diffuse emission or as an isotropic component from unresolved point sources. 
   The question of the origin of the extragalactic component cannot be answered
   without determining the galactic emission. 
   To discuss in detail all models resulting in gamma-ray background contributions 
   is far beyond the scope of this paper. Therefore the focus will be on recent publications on the 
   extragalactic high energy ($>$100~MeV) part of the gamma-ray background.
     }

   \authorrunning{T.M. Kneiske}            
   \titlerunning{Gamma-ray background: a review}  


   \maketitle
%
%
\section{Introduction}           
\label{sect:intro}
The first gamma photon ever detected by astronomers was a background photon
possibly of galactic origin. 
Since then observations have improved and a lot of ideas have been introduced 
to explain the data. It is still under debate what causes diffuse emission. 
Is it isotropic radiation due to particle processes in our Galaxy or in the 
universe or just a faint residue by gamma-ray sources which are too dim to be detected 
by recent telescopes and observatories.
The gamma-ray background signal is so important because it is 
a strict upper limit for theoretical models of possible contribution. 
For example, models including number density of the extragalactic contribution by unresolved point 
sources has to predict a gamma-ray flux which is below the observed signal. 
If gamma rays and neutrinos are produced in the same process, one can also derive 
an upper limit for the neutrino background using the extragalactic component of 
the diffuse gamma-ray observation.
The paper will focus on energies above 100~MeV and is organized as follows. A 
brief summary of observations will be followed by the question how to 
determine the galactic flux. The next section will focus 
on different contributions by unresolved point sources after showing a general 
method for the calculation, including cascade emission which is initiated by 
the annihilation of gamma-ray and low energy background photons. In the 
last chapter other possible contribution will be discussed.


\section{Observations}
\label{sect:Obs}
In 1965 Kraushaar published the first detection of a gamma-ray photon above 100 MeV. 
The total flux was $F=3 \cdot 10^{-4}$ cm$^{-2}$ s$^{-1}$ sr$^{-1}$ (Kraushaar et al. 1965) 
which is still a factor of 20 higher than the number of recent observations.
A few years later the same authors published results of the OSO-3 satellite which 
not only showed an isotropic component but also could distinguish between a strong 
galactic and a fainter extragalactic component (Kraushaar et al. 1972).
The first spectrum above 35~MeV could be derived with data from SAS-2 with a spectral 
index of -2.35 (Thomson \& Fichtel 1982). This number is very close to the -2.1 which 
can be fitted to the extragalactic data taken with EGRET (Sreekumar 1998). GLAST which to
be launched by the end of this or the beginning of next year will be able to improve 
the observation due to its much better resolution and sensitivity.
After detecting the total signal the problem 
of distinguishing of galactic and extragalactic components occurs.

\section{Galactic Emission}
\label{sect:gal}
The determination of galactic emission which is more than one order
of magnitude higher is crucial. 
The sky map has to be divided into several regions to get different flux levels.
The extragalactic component is assumed to be isotropic, which leads to the
same flux and the same spectrum for each region. A theoretical model has to be
developed for the galactic component (GB). A subtraction for each region with this
model should lead to the same residue, which is the extragalactic component.
The uncertainties can be reduced by fitting only the shape of the galactic flux
but not the absolute value. By introducing a normalization factor ($c_1$) for 
each sky region
the dependence on fluctuations in gas densities, in the interstellar radiation field and in the
cosmic ray densities can be reduced.
In our Galaxy the interactions of cosmic-ray protons with the interstellar gas
and electrons interacting with stellar photons are believed to be the main mechanisms of gamma-ray emission.
At MeV energies inverse Compton scattering
and bremsstrahlung are dominating the flux, while the GeV photons are produced in neutral
pion decay (Stecker 1977). A first analysis of Sreekumar et al. (1998)
led to a spectrum which could be fitted by the power law with the spectral index of $\alpha=-2.1$.
Using an updated galactic model, Strong, Moskalenko \& Reimer (2004) recalculated the
EGRET data and found a smaller flux with an excess above 1~GeV.


The idea that another galactic contribution 
could come from dark matter (DM) annihilation was presented by De Boer et al. (2005).
The excess in the EGRET data above 1~GeV is then explained by a dark matter annihilation
signal from Weakly Interacting Massive Particles (WIMPS) in a mass range from $\approx$
50 to 100~GeV. The resulting extragalactic gamma-ray background was published 
by De Boer et al. (2007) and is closer to the results of Strong, Moskalenko \& Reimer (2004).

This would lead to a formalism 
where for every given direction $\Xi$ in the sky the flux can be written as

\begin{equation}
F_{\mathrm obs}(\Xi) = c_1 \cdot F_{\mathrm GB} + F_{\mathrm EGB} + c_2 \cdot F_{\mathrm DM}
\end{equation}

while the dark matter contribution is still under discussion.
For example in a recent study Stecker, Hunter \& Kniffen (2007) were re-examining in detail the so called GeV "anomaly".
They found that instead of an astrophysical phenomenon it could be explained by 
correcting the flux sensitivity of EGRET above 1~GeV. Their analysis confirmed the results
by Sreekumar et al. (1998).

Another very detailed analysis of the background determination is published by Keshet, Waxman \& Loeb (2004).
They show that methods previously used to identify the Galactic emission 
depend on the Galactic tracers used and on the part of the sky examined.
In comparison to the other results they found a quite low flux at 1~GeV in their analysis.

Keeping the problems of data analysis and galactic emission 
in mind we will proceed with possible interpretations of the extragalactic component.

\section{Faint cosmic sources}
\subsection{Method}
To calculate the total flux contribution to the extragalactic background 
by a population of unresolved sources, the following equation is used

\begin{equation}
\frac{dN}{dE_\gamma\ d\Omega} =  \frac{1}{4\pi} \int^{z_m}_0  \frac{dV_c}{dz}
\int^{\infty}_{L_{\mathrm m}} \frac{dN}{dV\ dL} \frac{dN^i}{dE_\gamma}(z)\ \mathrm{d}L
 \ \mathrm{d}z,
 \ \ \ \
 \label{eq:gammabackOhne}
\end{equation}
with $dN^i/dE_\gamma(z)$ as the intrinsic gamma-ray flux. 
$dV_c/dz$ as the cosmological volume element, 
$L_{m}$ as the total luminosity of the weakest source and
the luminosity function $dN/(dV\ dL)$.

A template spectrum $dN/dE_\gamma $ can be derived by averaging observed gamma-ray 
spectra for a certain source population. If no observations are available, a theoretical 
model has to be developed based on average parameters derived from observations 
in other wavelengths.
Crucial for the calculation is the gamma-ray luminosity function $dN/(dL dV)$. 
If a statistical relevant number of sources have been detected in gamma rays 
a local luminosity function can be derived including a term for density and/or 
luminosity evolution. Without gamma-ray observations the luminosity function 
has to be calculated using observations in a different energy range and a correlation 
function. 

\subsection{Absorption and Cascade Emission}
Is the gamma-ray spectrum of an extragalactic source 
extending to energies above $\approx 20$~GeV, extragalactic 
absorption due to photon-photon pair production with low energy background photons 
has to be taken into account. The produced electron-positron pair is initiating an 
inverse Compton - pair cascade which leads to a gamma-ray flux by secondary photon production. 
For a simple analytical description equation~2 can be modified by taking absorption and the first
generation of secondary gamma-ray photons from cascade emission into account

\begin{eqnarray}
\frac{dN}{dE_\gamma\ d\Omega} & = & \frac{1}{4\pi} \int^{z_m}_0  \frac{dV_c}{dz}
\int^{\infty}_{L_{\mathrm m}} \frac{dN}{dV\ dL}  x \nonumber \\
& x & \left[\frac{dN^i}{dE_\gamma}(z)+\frac{dN^c}{dE_\gamma}(z)\right]\ 
e^{-\tau_{\gamma \gamma}(z)}\ \mathrm{d}L
 \ \mathrm{d}z_s,
 \ \ \ \
 \label{eq:gammaback}
\end{eqnarray}
with the cascade emission $dN^c/dE_\gamma(z, L)$
and pair creation optical depths $\tau_{\gamma\gamma}\gg1$.

For a detailed calculation a monte carlo cascade code has to be used.
The absorption is due to a photon background at ultraviolet, optical and infrared 
energies which is produced by stars in galaxies. The measurements of the so called 
extragalactic background light are leaving room for uncertainties within one order 
of magnitude. For other redshifts no direct observations can be obtained.
Models for the extragalactic background light (EBL) have been developed by several authors
(see the review by Hauser \& Dwek 2001). To calculate the absorption and cascade emission 
for a population of sources at low and high redshift 
the redshift evolution of the EBL has to be
taken into account (e.g. Salamon \& Stecker 1998, Kneiske et al. 2002, 2004).
The models include optically selected galaxies and infrared galaxies 
by spectral synthesis models, a cosmic star formation rate and the physics
of the interstellar medium. The result is the optical to infrared flux
as a function of redshift, where the optical part is due to direct starlight while 
the infrared emission is re-radiated starlight by interstellar dust. 
A model based on Kneiske et al. (2004) is shown for four selected redshifts in Fig.~1.
The EBL model takes new observations into account
by choosing values for model parameters as stated in figure 1.

   \begin{figure}
   \centering
   \includegraphics[width=\textwidth,height=120mm]{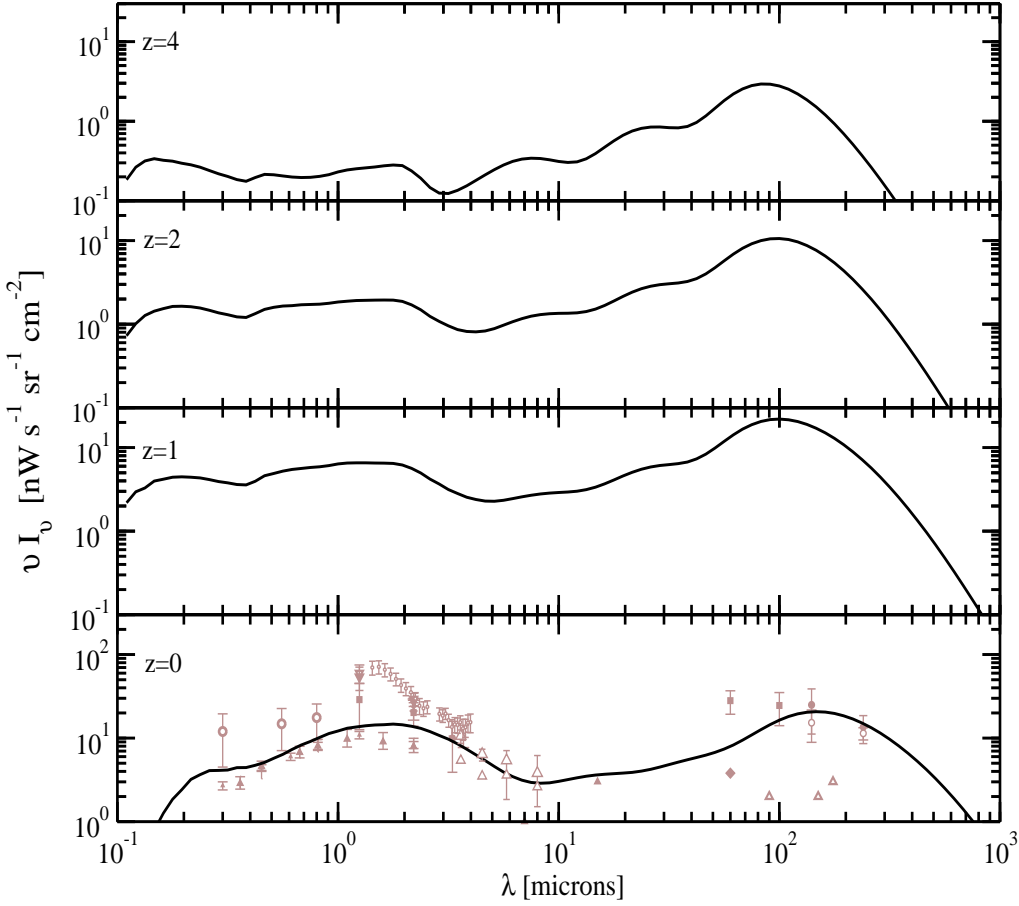}
   \caption{The metagalactic radiation field for four selected redshifts. 
   For data point references, a detailed model description and parameter definition see Kneiske et al. (2004, 2007).
   The values for the shown EBL model are: SFR$_{opt}$=(3.5, -1.2, 1.2, 0.1), SFR$_{LIG}$=(4.5, 0, 1.0, 0.1), 
   f$_{esc}$=0; c$_2=pow(10, -23.4)$. They are chosen to account for recent observations by the infrared
   satellite SPITZER. 
   }
   \label{Fig:MRF}
   \end{figure}

The electrons and positrons interact via inverse Compton scattering with 
cosmic microwave photons. The interaction takes place at a distance of a few hundred Mpc 
to a few Gpc depending on the energy of the primary photon (Protheroe \& Stanev 1993). This is outside of the 
large scale cosmic structure where the magnetic field is assumed to be very small. 
Depending on the actual strength of the magnetic field the secondary gamma-rays 
are beamed into the same direction as the primaries and would simply add to the 
primary flux. In case of a high magnetic field, they are distributed isotropically 
in a halo around the gamma-ray source.

\subsection{Star Forming Galaxies}
Based on the fact that high energy gamma-ray emission has been observed in our own
Galaxy a so called guaranteed flux from other galaxies can be calculated. 
Pavlidou \& Fields (2002) used a power law fit of the Galactic spectrum with a spectral
break at 850~MeV, a total gas mass of $10^{10}$~M$_{\odot}$, a gas density of 1~cm$^{-3}$
and a star formation of 3.2~M$_\odot$ yr$^{-1}$ to get 
luminosity at gamma-ray energies for a Milky-Way-like galaxy.
Integrating the luminosity with the cosmic star formation rate up to a redshift
of 5, they got a flux of 2 and 6 x $10^{-7}$~GeV cm$^{-2}$ s$^{-1}$ sr$^{-1}$
for a star formation without and with dust correction respectively.
Most of the flux is emitted by galaxies with redshifts smaller than $z=1$.

The same authors have calculated contribution by unidentified EGRET sources (Pavlidou et al. 2007).
The idea is, if these sources are of extragalactic origin they contribute as a distinct
extragalactic population. If these sources are located within the Galaxy, their counterparts
in external galaxies could  contribute to the unresolved emission from theses systems.
For this calculation two assumptions had to be made.
The first is that all unidentified EGRET sources without counterpart are in fact
one class of objects, so that a flux distribution can be obtained. 
The second assumption is that the flux distribution can be extrapolated to 
the faint end.
The result is in agreement within the confident limits of the EGRET background
data by Strong et al. (2004). This is quite interesting although
it is questionable that unidentified EGRET sources can be 
taken as distinct class of objects.

A more extreme class of star forming galaxies are the starburst and luminous infrared galaxies (LIG).
They have high infrared luminosities, high gas densities and star formation rates 
which are a factor of ten higher than in the Milky Way. 
Therefore they are very good candidates for high energy gamma-ray emission although
it has not been detected yet.
Based on the assumption that relativistic protons lose nearly all their energy due to 
pion production within a luminous infrared galaxy and that the observed radio
synchrotron flux is only emitted by secondary electrons Thompson, Quataert \& Waxman (2006)
came to a 10\% contribution of the extragalactic gamma-ray background.
In their calculation some of the favorable  LIGs and starbursts (M82, NGC253, and IC342) 
should be observable with the next 
gamma-ray satellite GLAST.
This result was questioned by Stecker (2006). He argued that even if the assumptions
were correct, another problem occurred. The gamma-ray background calculation is
normalized to the total local infrared luminosity density from the IRAS2Jy sample
(Yun et al. 2001).
From the radio-FIR relation, which seems to be also valid for starburst regions, a
radio luminosity function at 1.4~GHz is calculated. And because of the assumption that 
the radio and the gamma-ray flux are both produced in pion decay, it is straightforward 
to derive a gamma-ray luminosity function.
Stecker pointed out that not 100\% of the local infrared luminosity density is due
to emission from starbursts but rather 10\%. Including the emission from starburst at higher redshift
the percentage comes to about 23\%. Therefore the contribution of starburst galaxies is
a factor of five smaller.

\subsection{Active Galactic Nuclei}
Back in 1996 Stecker \& Salamon calculated a contribution of blazars using a radio 
luminosity function and a linear correlation between radio and gamma-ray luminosity.
They also included a flaring component of blazars taking the steepening of the spectrum
into account. The result was that blazars could account for 100\% of the observed flux.
They added the effect of extragalactic absorption in Salamon \& Stecker (1998).
Chiang \& Mukherjee (1998) showed that Stecker \& Salamon had
failed to reproduce the observed number of sources detected by EGRET 
by overproducing the number at low redshift.
Stecker (2001) argued that his assumption of a correlated radio and gamma-ray emission in blazars
is contrary to the statistical independent analysis in CM98 which introduces a bias.
Thus, the calculation in Chiang \& Mukherjee (1998) using the EGRET luminosity function and a radio luminosity
function for completeness came only to 25\% - 50\% of the observed background flux.
A simple correlation between luminosities is always a source of uncertainty so other
models started with multiwavelength spectra.
Giommi et al. (2005) modeled a synchrotron self Compton spectrum for blazars.
The spectrum was normalized to radio and X-ray data. Using a radio luminosity function
they could reproduce the X-ray background very well but had problems at higher energies.
Other models by Muecke \& Pohl (2000) included a contribution of BL Lacs and Flat Spectrum
Radio Quasars separately. From this calculation 40\%-80\% of the background are due to unresolved AGN.
This work was updated recently by Dermer (2007) where he used a physical model to fit the 
redshift and size distribution of EGRET blazars.

A contribution of secondary gamma-rays from BL Lacs has been calculated by Kneiske \& Mannheim (2007).
Direct emission of X-ray BL Lac (XBL) can only make a minor contribution, since the
maximum of high energy emission is at TeV energies. But the secondary gamma-rays have energies 
about two orders of magnitude below the primary photons.
An X-ray luminosity function showing almost no evolution has been used.
The result was a contribution of about 10\% to the observed GeV background which is almost
all due to secondary flux. 

A similar calculation has been done by Stawarz, Kneiske \& Kataoka (2006) for the extended
jet emission from Faranoff-Riley (FR) galaxies. Assuming that the observed X-ray emission
in the bright knots of the kiloparsec scale jets are due to synchrotron emission, a
gamma-ray flux can be calculated. Including the secondary gamma-ray photons we found that only
1\% of the gamma-ray background could be explained by the extended luminosity of FRI galaxies.

   \begin{figure}
   \centering
   \includegraphics[width=\textwidth,height=120mm]{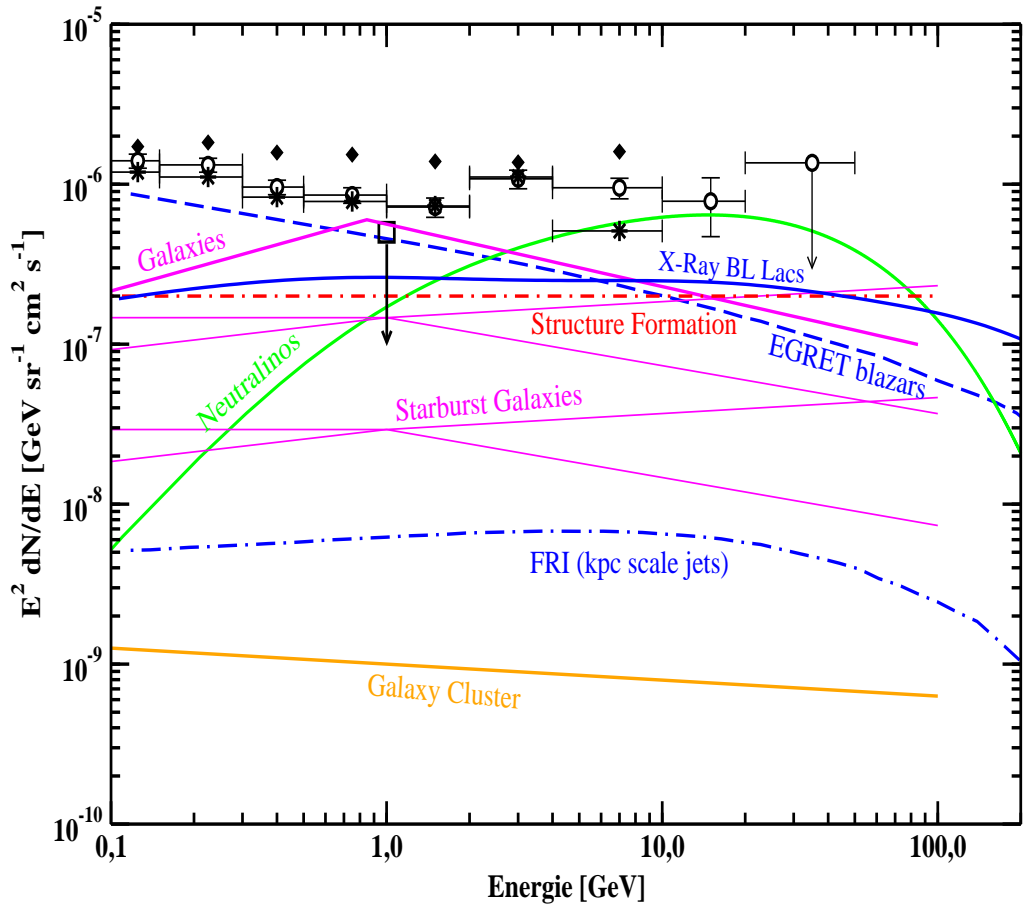}
   \caption{Gamma-ray background contributions. Shown are contribution by
   normal galaxies (thick pink line; Pavlidou \& Fields 2002);
   starburst galaxies (thin pink line; Thompson, Quataert \& Waxman 2007);
   EGRET blazars (blue dashed line; Kneiske \& Mannheim 2007);
   X-Ray BL Lacs (blue solid line; Kneiske \& Mannehim 2007);
   Neutralinos (green solid line; Elsaesser \& Mannheim 2005);
   FRI (kpc scale jets (blue dot-dashed line; Stawarz, Kneiske \& Kataoka 2006);
   Galaxy Clusters solid orange line; Colafrancesco \& Blasi 1998);
   Structure Formation (red dot-dashed line; Miniati 2002).
   }
   \label{Fig:GRB}
   \end{figure}

\subsection{Gamma-Ray Bursts}

In the analysis by Casanova, Dingus \& Zhang (2007) the contribution of gamma-ray bursts 
have been calculated. A power-law for the synchrotron and inverse Compton component
has been used where the inverse Compton flux is higher by a factor of ten.
Extragalactic absorption has been taken into account too.
The result is a 10\% contribution at GeV energies.

\subsection{Galaxy clusters}
Galaxy clusters have not been detected by EGRET, but an analysis by
Scharf \& Mukherjee (2002) has shown that a possible correlation exists between
high Galactic latitude EGRET data and Abell clusters ($\ge 3~\sigma$).
They have shown that 447 of the richest clusters with a bolometric luminosity
of $L~\approx 10^{44}$ erg s$^{-1}$ and no evolution could explain about 1\% to 10\%
of the gamma-ray background.

A more theoretical calculation by Colafrancesco \& Blasi (1998) is based on a self-consistent
picture of cluster formation and evolution.
The model starts from a primordial density perturbation spectrum, and a realistic modeling
for the distribution of the intergalactic medium. They found that an evolving population
of clusters can only produce up to 2\% of the observed flux.

\section{Other contributions}
Other ideas have been published on the origin of the extragalactic
gamma-ray background, like matter-antimatter annihilation 
(e.g. Stecker, Morgan \& Bredekamp 1971, Cohen, Rujula \& Glashow 1998)
or the decaying of primordial black holes (MacGibbon \& Carr 1991).
Purely diffuse orgins were discussed in Stecker 1973.
The resulting spectra are quite different
from what has been observed with EGRET, so it is very unlikely 
to have a significant contribution from this processes.
Some of the more recent results are the following.

Elsaesser \& Mannheim (2005) used high resolution simulations of structure
formation to calculate contribution with a maximum around 10~GeV 
from neutralino annihilation in cold dark matter halos. They found a neutralino
mass of 515~GeV for their best-fit model.

Gravitational induced shock waves produced during cluster mergers and large-scale structure
formation give rise to highly relativistic electrons that are responsible for
inverse Compton scattering of the cosmic microwave background photons to GeV energies.
A contribution to the gamma-ray background is produced in filaments, sheets, 
and extended gamma-ray halos associated with massive cluster (Loeb \& Waxman 2000).
Similar to this Miniati (2002) found that cosmic rays of cosmological origin can account
for about 20\% of the gamma-ray background. In his calculation 30\% of the computed flux
is emitted by the decay of neutral pions generated in p-p collisions of the ionic cosmic-ray
component with the thermal gas.

In Dado, Dar \& Rujula (2007) the authors explain the extragalactic gamma-ray background
by inverse Compton scattering of the cosmic microwave background and stellar photons by
cosmic-ray electrons in the interstellar and intergalactic space. In their work they 
get a much higher galactic contribution from cosmic-ray electrons in the Galactic halo.
The extragalactic emission is calculated from electrons ejected by supernova explosions
and AGN. 

A fluctuation analysis could give 
a better understanding which of the many possible origins are dominating.
The angular power spectrum of intensity fluctuations 
of the extragalactic gamma-ray background measured in the future by GLAST could 
probe its origin (Miniati et al. 2007).

\label{lastpage}

\end{document}